\documentclass[a4paper]{revtex4}
\usepackage{graphicx}
\usepackage{fancyhdr}
\usepackage{amsmath}
\usepackage{color}
\usepackage{booktabs}
\usepackage{caption}

\newcommand{\IE}{\textit{i.\,e.}}
\newcommand{\I}{{i\mkern1mu}}
\newcommand{\E}{\mathrm{e}}
\def\unity{{\bf 1}}

\begin{document}

\title{Top Yukawa-Coupling Enhanced Two-Loop Corrections to the Masses of the Higgs Bosons in the MSSM with CP Violation}

%

\author{Wolfgang Hollik}
\affiliation{Max-Planck-Institut f\"ur Physik (Werner-Heisenberg-Institut), F\"ohringer Ring 6, D--80805 M\"unchen, Germany}

\author{Sebastian Pa{\ss}ehr}
\affiliation{Deutsches Elektronen-Synchrotron DESY,\\ Notkestra{\ss}e 85, D--22607 Hamburg, Germany}

\begin{abstract}
Recent results for the leading $\mathcal{O}{\left(\alpha_t^2\right)}$ two-loop corrections
to the Higgs-boson masses of the MSSM with complex parameters are presented.
The strategy of the Feynman-diagrammatic, analytical calculation is illustrated.
Numerical analyses show agreement with a previous result in the limit of real parameters.
Furthermore, the newly available dependence on complex MSSM parameters is investigated.
Mass shifts by \mbox{$\approx 1$}GeV for different phases underline the importance of this contribution
for a precise prediction of the Higgs-boson mass spectrum in the MSSM.
\end{abstract}

\maketitle

\thispagestyle{fancy}


\section{Introduction}

The Higgs-like particle which has been discovered by the experiments
ATLAS~\cite{Aad:2012tfa} and CMS~\cite{Chatrchyan:2012ufa} at the LHC has
given rise to substantial investigations to reveal the nature of this
particle as a Higgs boson responsible for electroweak symmetry
breaking. Besides the presently consistent explanation of the
measurements by the Standard Model Higgs
boson~\cite{Landsberg:2013ina}, a large variety of other
interpretations is possible where the Higgs particle belongs to an
extended model connected to physics beyond the Standard Model.
The observed particle could also be explained as a light state within
a richer spectrum of scalar particles as predicted by the
theoretically well motivated minimal supersymmetric Standard
Model~(MSSM). The Higgs sector of the MSSM consists of two complex
scalar doublets leading to five physical Higgs bosons ($CP$-even
$h,H$, $CP$-odd $A$, charged $H^{\pm}$) and three (would-be) Goldstone
bosons. At the tree level their masses can be parametrized in terms of
the $A$-boson mass $m_A$ and the ratio of the two vacuum expectation
values, $\tan\beta = \left.v_2\middle/v_1\right.$. $CP$-violation is
induced in the Higgs sector via loop contributions involving complex
parameters from other SUSY sectors leading to mixing between $h,\,H$
and $A$ in the mass eigenstates~\cite{Pilaftsis:1998pe}.
 
Masses and mixings in the neutral sector are strongly affected by loop
contributions. A lot of work has been invested into higher-order
calculations of the mass spectrum for the case of the MSSM with real
parameters~\cite{Heinemeyer:1998jw,Heinemeyer:1998np,Heinemeyer:1999be,Heinemeyer:2004xw,Borowka:2014wla,mhiggsFD3l,Zhang:1998bm,Espinosa:2000df,Brignole:2001jy,Casas:1994us,Degrassi:2002fi,Heinemeyer:2004gx,Allanach:2004rh,Martin:2001vx}
as well as complex
parameters~\cite{Demir:1999hj,Pilaftsis:1999qt,Carena:2000yi,Heinemeyer:2007aq,Frank:2006yh}.
The largest loop contributions originate from the top Yukawa
coupling~$h_t$,
or~\mbox{$\alpha_t=\left.h_t^2\middle/(4\pi)\right.$}. The class of
leading two-loop Yukawa-type corrections of
$\mathcal{O}{\left(\alpha_{t}^{2}\right)}$ has been calculated for the
case of real parameters~\cite{Espinosa:2000df,Brignole:2001jy},
applying the method of the effective potential. Together with the full
one-loop result~\cite{Frank:2006yh} and the leading
$\mathcal{O}{\left(\alpha_{t}\alpha_{s}\right)}$
terms~\cite{Heinemeyer:2007aq}, both accomplished in the
Feynman-diagrammatic approach including complex parameters, it has
been implemented in the public program {\tt
  FeynHiggs}~\cite{Heinemeyer:1998np,Degrassi:2002fi,Frank:2006yh,Heinemeyer:1998yj,Hahn:2010te}.
The calculation of the $\mathcal{O}{\left(\alpha_{t}^{2}\right)}$
terms for the complex version of the MSSM, also performed in the Feynman-diagrammatic
approach, has been published
recently~\cite{Hollik:2014wea,Hollik:2014bua}. An outline of the calculation
and sample results for the lightest Higgs-boson mass are presented in this talk.

%

\section{Higgs masses in the complex MSSM}

The Higgs potential of the complex MSSM is given by
\begin{align}
  V_{\rm H} &= \frac{g_1^2 + g_2^2}{8}\left(\mathcal{H}_{2}^{\dagger}\mathcal{H}_{2} - \mathcal{H}_{1}^{\dagger}\mathcal{H}_{1}\right)^{2} + \frac{g_2^2}{2}\left|\mathcal{H}_{1}^{\dagger}\mathcal{H}_{2}\right|^2 + \sum\limits_{i\;=\;1}^2{\left(\tilde{m}_{i}^{2} + \lvert\mu\rvert^{2}\right)\,\mathcal{H}_{i}^{\dagger} \mathcal{H}_{i}} + \left(m_{12}^2\,\mathcal{H}_{1} \cdot \mathcal{H}_{2} + \text{h.\,c.}\right)
\end{align}
with the gauge couplings~$g_1,\, g_2$, the bilinear superpotential
parameter~$\mu$, and the soft-breaking parameters~$\tilde{m}_{i}$
(real) and~$m_{12}^{2} \equiv b_{\mathcal{H}_{1}\mathcal{H}_{2}}\,\mu$
(complex). The Higgs doublets are conventionally written in terms of
their charged and neutral components in the following way,
\begin{alignat}{2}
  \label{eq:Higgsfields}
  \mathcal{H}_{1} &= \begin{pmatrix} v_{1} + \frac{1}{\sqrt{2}}(\phi_{1} - \I \chi_{1})\\ -\phi^{-}_{1}\end{pmatrix},&\quad
  \mathcal{H}_{2} &= \E^{\I \xi}\begin{pmatrix} \phi^{+}_{2}\\ v_{2} + \frac{1}{\sqrt{2}}(\phi_{2} + \I \chi_{2})\end{pmatrix} .
\end{alignat}
Using the notation 
\mbox{$\Phi = \begin{pmatrix} \phi_{1}, & \phi_{2},  & \chi_{1}, & \chi_{2} \end{pmatrix}$}, 
\mbox{$\Phi^- = \begin{pmatrix} \phi^{-}_{1}, & \left(\phi^{+}_{2}\right)^{\dagger}\end{pmatrix}$}, 
\mbox{$\Phi^+ = \left(\Phi^-\right)^\dagger $},
\mbox{$\mathbf{M}_{\Phi}^{(0)} = \left(\begin{smallmatrix}\mathbf{M}_{\phi} & \mathbf{M}_{\phi\chi}  \\ 
               \mathbf{M}_{\phi\chi}^{\dagger} & \mathbf{M}_{\chi} \end{smallmatrix}\right)$},
the Higgs potential can be written as a polynomial in the field components,
\begin{align}
\label{eq:potential}
  \begin{split}
    V_{\rm H} &= -T_{\phi_{1}}\,\phi_{1} - T_{\phi_{2}}\,\phi_{2} - T_{\chi_{1}}\,\chi_{1} - T_{\chi_{2}}\,\chi_{2}
            + \frac{1}{2}\Phi\,\mathbf{M}_{\Phi}^{(0)}\,\Phi^{\dagger}
            + \Phi^{-}\,\mathbf{M}_{\Phi^{\pm}}^{(0)}\,\Phi^{+} + \dots  \ ,
  \end{split}
\end{align}
At the tree-level, the phases~$\xi$ and~$\arg{m_{12}^{2}}$ can be chosen as zero, 
and the $CP$-mixing entries~$\mathbf{M}_{\phi\chi}$ vanish.
Explicit expressions for the tadpole coefficients $T_i$ and for the mass matrices $\mathbf{M}^{(0)}$ 
are listed in Refs.~\cite{Frank:2006yh,Hollik:2014bua}.

The neutral Higgs-boson masses are derived from the poles of the propagator matrix~$\Delta_{\Phi}$, 
\begin{align}
\Delta_{\Phi} (p^2) &= - \Big[ \Gamma_{\Phi}(p^2) \Big]^{-1} \, ,
\quad {\rm with} \quad
\Gamma_{\Phi}(p^2)  \, = \,\Big[ p^2 \unity - \mathbf{M}_{\Phi} \Big] .
\end{align}
The irreducible two-point vertex functions~$\Gamma_{\Phi}$ contain
the mass matrix~$\mathbf{M}_{\Phi}$, which at lowest order is given by 
the constant matrix~$\mathbf{M}_{\Phi}^{(0)}$ 
in~Eq.~(\ref{eq:potential}).
At higher orders, $\mathbf{M}_{\Phi}$ 
is shifted by the momentum dependent matrix of the renormalized 
self-energies~$\mathbf{\hat{\Sigma}}_{\Phi}$ according to
\begin{align}
 \mathbf{M}_{\Phi} &\quad \to \quad \mathbf{M}_{\Phi}^{(0)} - \mathbf{\hat{\Sigma}}_{\Phi}(p^2)\ .
\end{align}
The self-energies contain in general mixing of all fields with equal quantum numbers.

In our case, we evaluate the momentum-dependent neutral ``mass matrix'' in the basis of tree-level mass eigenstates 
(\IE~$\{\phi_1,\,\phi_2,\,\chi_1,\,\chi_2\}\to \{h,\,H,\,A,\,G\}$) 
perturbatively up to the two-loop level,
%
\begin{align}
  \label{eq:masscorr}
    \mathbf{M}_{ h H A G}(p^2) &= 
    \mathbf{M}_{ h H A G}^{(0)} - \mathbf{\hat{\Sigma}}_{h H A G}^{(1)}(p^2) - \mathbf{\hat{\Sigma}}_{h H A G}^{(2)}(0) \ .
\end{align}
Therein, $\mathbf{\hat{\Sigma}}_{h H A G}^{(k)}$ denotes the matrix of
the renormalized diagonal and non-diagonal self-energies for the $h,
H, A, G$ fields at loop order $k$.  For the complex MSSM, the one-loop
self-energies are completely known~\cite{Frank:2006yh}. The
leading two-loop $\mathcal{O}{\left(\alpha_t \alpha_s\right)}$
contributions have been obtained in the approximation of zero external
momentum ~\cite{Heinemeyer:2007aq}, and the same approximation 
has been applied to derive the leading
Yukawa contributions of~$\mathcal{O}{\left(\alpha_t^2\right)}$,
described in detail in Ref.~\cite{Hollik:2014bua}, 
extending the on-shell renormalization scheme of Ref.~\cite{Frank:2006yh} 
to the two-loop level.

The masses of the three neutral Higgs bosons
including the new $\mathcal{O}{\left(\alpha_{t}^{2}\right)}$
contributions are given by the real parts of the poles of the
$hHA$-propagator matrix, obtained as the zeroes of the determinant of
the renormalized two-point vertex function, {\it i.e.} solving
\begin{align}
  \label{eq:higgspoles}
   \operatorname{det}  
     \left[p^2 {\unity} - \mathbf{M}_{hHA}{\left(p^2\right)}\right] 
 & = \, 0 \, ,
\end{align}
with the corresponding $(3\times 3)$-submatrix of
Eq.~\eqref{eq:masscorr}. Mixing with the Goldstone boson yields
subleading two-loop contributions; also Goldstone--$Z$ mixing occurs
in principle, which is related to the other Goldstone mixings by
Slavnov--Taylor identities~\cite{Baro:2008bg,Williams:2011bu} and of
subleading type as well~\cite{Hollik:2002mv}.  However, mixing with
Goldstone bosons has to be taken into account inside the loop diagrams
and for a consistent renormalization. 

The renormalized self-energies in Eq.~\eqref{eq:masscorr} require  
counterterms up to second order in the loop expansion. 
A detailed study of all required counterms as well as
explicit expressions for the renormalization constants are given in
Ref.~\cite{Hollik:2014bua}.

\section{Numerical Results}

In this section numerical analyses for the masses of the neutral Higgs
bosons derived from Eq.~\eqref{eq:higgspoles} are presented. The
complete one-loop result including the dependence on the external
momentum, and the $\mathcal{O}{\left(\alpha_{t}\alpha_{s}\right)}$
terms are obtained from {\tt FeynHiggs}, while the
$\mathcal{O}{\left(\alpha_{t}^{2}\right)}$ terms are computed by means
of the corresponding two-loop self-energies specified in the previous
section. The combination of all contributions is carried out according
to Eq.~\eqref{eq:masscorr} within {\tt FeynHiggs}.

The~SM~parameters are put together in Tab.~\ref{tab:parameters}, as
well as those~MSSM~parameters that are kept for the following
analyses. The residual input parameters of the~MSSM are shown in the
figures or their captions.  The parameters~$\mu,\, t_{\beta}$ and the
Higgs field-renormalization constants are defined in
the~$\overline{\text{DR}}$~scheme at the scale~$m_{t}$.

\begin{table}[h]
  \centering
  \caption[Default input parameters]{Default input values of the MSSM and SM parameters.}
  \label{tab:parameters}
  \begin{tabular}{r@{$\;=\;$}lr@{$\;=\;$}l}
    \hline
    \multicolumn{2}{c}{MSSM input} & \multicolumn{2}{c}{SM input}\\
    \hline
    $M_2$ & $200$~GeV, & $m_t$ & $173.2$~GeV,\\
    $M_1$ & $\left.\left(5s_{\rm w}^2\right)\middle/\left(3c_{\rm w}^2\right) M_2\right.$, & $m_b$ & $4.2$~GeV,\\
    $m_{\tilde{l}_1} = m_{\tilde{e}_{\rm R}}$ & $2000$~GeV, & $m_{\tau}$ & $1.77703$~GeV,\\
    $m_{\tilde{q}_1} = m_{\tilde{u}_{\rm R}} = m_{\tilde{d}_{\rm R}}$ & $2000$~GeV, & \qquad$M_W$ & $80.385$~GeV,\\
    $A_u = A_d = A_e$ & $0$~GeV, & $M_Z$ & $91.1876$~GeV,\\
    $m_{\tilde{l}_2} = m_{\tilde{\mu}_{\rm R}}$ & $2000$~GeV, & $G_{\text{F}}$ & $1.16639\cdot 10^{-5}$,\\
    $m_{\tilde{q}_2} = m_{\tilde{c}_{\rm R}} = m_{\tilde{s}_{\rm R}}$ & $2000$~GeV, & $\alpha_s$ & $0.118$.\\
    $A_c = A_s = A_\mu$ & $0$~GeV,\\
    \hline
  \end{tabular}
\end{table}

\begin{minipage}{.49\textwidth}
  \centering
  \includegraphics[width=1.05\textwidth]{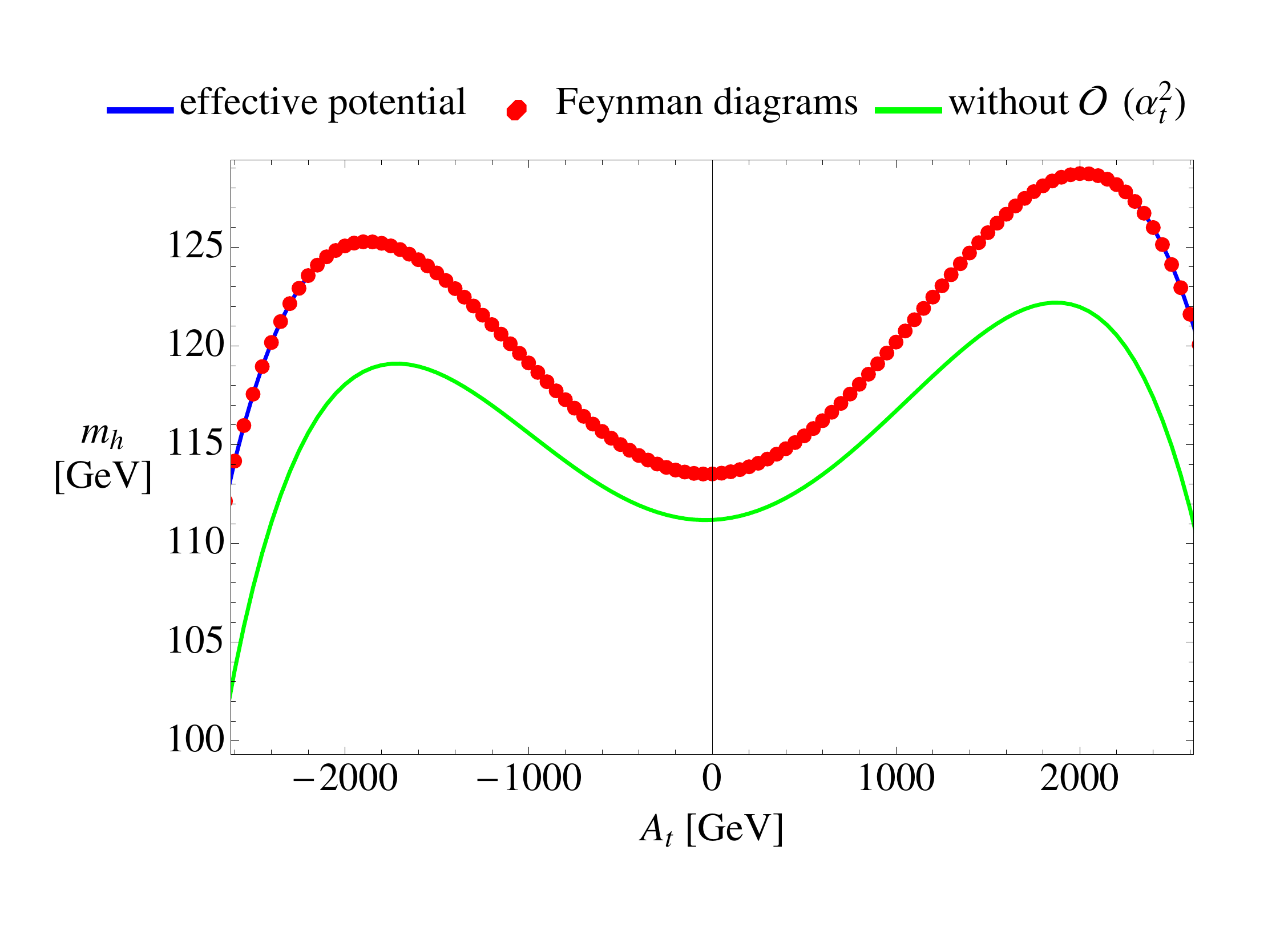}
  \quad\\[-3ex]
  \captionof{figure}{\label{fig:realAtcompare}Comparison of the
    lightest Higgs-boson mass in the effective potential approach
    (blue) and the Feynman-diagrammatic approach (red). The curves are
    lying on top of each other, indicating the agreement of both
    calculations in the limit of real parameters. For reference the
    result without the contributions of
    $\mathcal{O}{\left(\alpha_{t}^{2}\right)}$ is shown (yellow). The
    input parameters are~\mbox{$m_A = 800$~GeV}, \mbox{$\mu = 200$},
    \mbox{$t_{\beta} = 30$}, \mbox{$m_{\tilde{q}_3} =
      m_{\tilde{t}_{\text{R}}} = m_{\tilde{b}_{\text{R}}} =
      1000$~GeV}, \mbox{$m_{\tilde{\ell}_3} = m_{\tilde{\tau}_{\rm R}}
      = 1000$~GeV}, \mbox{$m_{\tilde{g}} = 1500$~GeV}, \mbox{$A_t =
      A_b = A_{\tau}$}.}
\end{minipage}\quad
\begin{minipage}{.49\textwidth}
  \quad\\[-7ex]
  \centering
  \includegraphics[width=.95\textwidth]{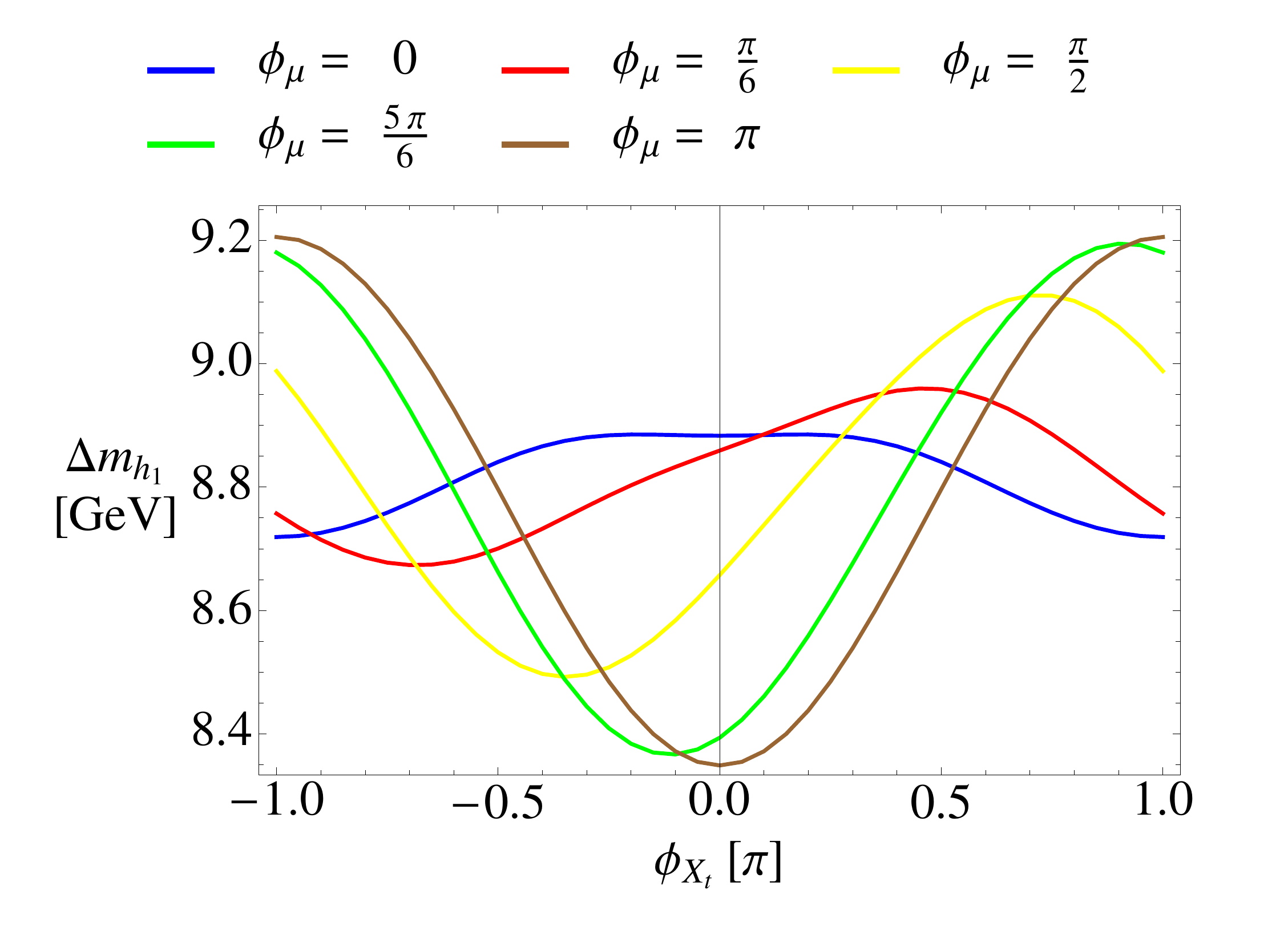}
  \captionof{figure}{\label{fig:Mh1complexshift}%
    The shift of the lightest Higgs-boson mass by the phases~$\phi_{X_t}$ and~$\phi_\mu$ with respect to the case of zero phases, i.\,e.~\mbox{$\Delta m_{h_1} = m_{h_1}\left(\phi_{X_t},\,\phi_\mu\right) - m_{h_1}\left(\phi_{X_t} = 0,\,\phi_\mu = 0\right)$}. The input parameters are~\mbox{$m_{H^\pm} = 200$~GeV}, \mbox{$\lvert\mu\rvert = 2500$~GeV}, \mbox{$t_{\beta} = 10$}, \mbox{$m_{\tilde{\ell}_3} = m_{\tilde{\tau}_{\rm R}} = 1000$~GeV}, \mbox{$m_{\tilde{q}_3} = m_{\tilde{t}_{\text{R}}} = m_{\tilde{b}_{\text{R}}} = 1500$~GeV}, \mbox{$m_{\tilde{g}} = 2000$~GeV}, \mbox{$\lvert X_t \rvert = 2\,m_{\tilde{q}_3}$}, \mbox{$A_b = A_{\tau} = 0$}.}
\end{minipage}

\medskip
A comparison of the obtained result in the real MSSM with the
previously known $\mathcal{O}{\left(\alpha_{t}^{2}\right)}$
contributions~\cite{Brignole:2001jy} from a calculation making use of
the effective-potential method for the mass of the lightest Higgs
boson has been presented recently in Ref.~\cite{Hollik:2014wea}. An
example which shows very good agreement is depicted in
Fig.~\ref{fig:realAtcompare}.

The phase depending contribution to the lightest Higgs mass~$\Delta
m_{h_1}$ is illustrated in Fig.~\ref{fig:Mh1complexshift} showing a
possible mass shift of~$\approx 1$~GeV for different choices
of~$\phi_{X_t}$ (with~\mbox{$X_t = A_t^* -
  \left.\mu\middle/t_\beta\right.$}) and~$\phi_\mu$.

\section{Conclusions}

We have presented the leading
$\mathcal{O}{\left(\alpha_{t}^{2}\right)}$ corrections to the
Higgs-boson masses of the MSSM with complex parameters. In the limit
of real parameters a previous result is confirmed. Combining the new
terms with the existing one-loop result and leading two-loop terms of
$\mathcal{O}{\left(\alpha_t\alpha_s\right)}$ yields an improved
prediction for the Higgs-boson mass spectrum also for complex
parameters that is equivalent in accuracy to that of the real
MSSM.

The numerical discussion illustrates that the mass shifts originating
from the $\mathcal{O}{\left(\alpha_{t}^{2}\right)}$ terms are
significant, and hence an adequate treatment also for complex
parameters is an obvious requirement. Besides the mass shift of
approximately~$5$~GeV in the real~MSSM a further shift of~$\approx
1$~GeV can be induced by complex parameters.

The new terms will be included in the publicly available code {\tt
  FeynHiggs}.

\begin{acknowledgments}
This work has been supported by the Collaborative Research Center
SFB676 of the DFG, "Particles, Strings and the early Universe".
\end{acknowledgments}

\bigskip 

\end{document}